\documentclass[twocolumn]{article}

\usepackage{graphicx}
\usepackage{amsmath,amssymb}
\usepackage{float}
\usepackage{algorithm}
\usepackage{algpseudocode}
\usepackage{braket}
\usepackage{xurl}
\usepackage{array}

\usepackage{natbib}

\providecommand{\textcite}{\citet}
\providecommand{\parencite}{\citep}

\begin{document}
\makeatletter
\let\trimmarks\relax
\makeatother

\markboth{Muhammad Faryad et al.}{Hybrid Quantum--Classical k-Means Clustering via Quantum Feature Maps}



\title{Hybrid Quantum--Classical k-Means Clustering via Quantum Feature Maps}

\author{Syed M. Abdullah, Alisha Baba, Muhammad Siddique\\
Department of Computer Science, Lahore University of Management Sciences, Lahore 54792, Pakistan\\
Muhammad Faryad
\\Department of Physics, Lahore University of Management Sciences, Lahore 54792, Pakistan\\
muhammad.faryad@lums.edu.pk}

\maketitle

\begin{abstract}
Clustering is one of the most fundamental tasks in machine learning, and the k-means clustering algorithm is perhaps one of the most widely used clustering algorithms. However, it suffers from several limitations, such as sensitivity to centroid initialization, difficulty capturing non-linear structure, and poor performance in high-dimensional spaces. Recent work has proposed improved initialization strategies and quantum-assisted distance computation, but the similarity metric itself has largely remained classical. In this study, we propose a quantum-enhanced variant of k-means that replaces the Euclidean distance with a quantum kernel derived from the inner product between feature-mapped quantum states. Using the Iris dataset, we use multiple quantum feature maps—including entangled SU2 and ZZ circuits—to embed classical data into a higher-dimensional Hilbert space where cluster structures become more separable. We will also be testing using another dataset, namely the breast cancer dataset. Similarity between data points is computed through the inner product between two states. Our results show that this approach achieves improved clustering stability and competitive accuracy compared to the classical algorithm, with the SU2 feature map yielding an accuracy of 88.6\% on the Iris dataset and 91.0\% on the breast cancer dataset, despite operating on NISQ-feasible shallow circuits. These findings suggest that quantum kernels provide a richer similarity landscape than traditional distance metrics, offering a promising path toward more robust unsupervised learning in the NISQ era.
\end{abstract}

\section{Introduction}
Unsupervised learning is a branch of machine learning that deals with training models on unlabeled data. Such algorithms explore the dataset, look for similarities between data points, and group these points into meaningful clusters. Unlike supervised learning, there is a natural grouping without any guidance. A commonly used technique is k-means clustering, which partitions data into k groups by repeatedly assigning each point to the nearest cluster centre, called the centroid, using distance as a metric and updating those centroids until the data point assignments become stable. The distance could be any, for example, the Manhattan or Euclidean distance. The traditional k-means clustering algorithm has two major weaknesses:
\begin{itemize}
    \item The centroids are chosen randomly at initialization
    \item When the dataset is high-dimensional, you get stuck in a local minima of the cost function
\end{itemize}
\cite{li2012} Zhang and Li in their paper, proposed an improved version of the classical \textit{k}-means algorithm, where instead of a random choice of centroids during initialization, you select the centroids that are as far apart as possible from each other. This leads to faster convergence and higher clustering precision. The experimental results showed a lower error value and fewer iterations compared to standard \textit{k-means}, confirming that better initialization enhances both accuracy and stability.
Similarly, \cite{zubair2024} M. Zubair addresses the issue of inefficient initial centroid selection in \textit{k-means} by proposing Principal Component Analysis (PCA) and percentile-based centroid selection. This ensures that centroids are spread out across the dataset, thereby improving both convergence speed and clustering accuracy. Their results show that this reduces the execution time as compared to the traditional algorithm. These two papers serve as motivation for our project, in which we aim to enhance the \textit{k-means} algorithm by integrating it with quantum computing. We hope to improve the algorithm by using a quantum feature map, which utilizes parameterized quantum circuits to define kernels that can naturally separate data before clustering. Then, instead of using distance as a metric, we calculate the inner product between quantum states. This works because the inner product measures similarity between states, much like how distance measures similarity in traditional methods. Thus, while Li, Wu, and Zubair improved initialization in the spatial domain, our approach seeks similar stability and precision gains through quantum computing.
\newline

There have also been efforts on the Quantum side to improve k-means clustering. One such effort belongs to Poggiali
et al. \cite{poggiali2024} and their team, who acknowledge that computing Euclidean distance for each data point in a large dataset can become a bottleneck for the algorithm, making it extremely slow. To demonstrate this, the authors suggested a hybrid quantum-classical model using quantum circuits to speed up distance computation while still keeping the structure of the classical k-means algorithm. They used different variants, for example, the Swap Test, to compute all distances from a point to multiple centroids simultaneously. Their results showed a speed-up, increasing the algorithmic efficiency and hence optimizing the k-means algorithm. While their focus was on efficiency through quantum parallelism, our paper hopes to use quantum feature maps to alter the geometry of the data and improve cluster separation.
However, we will see how choosing the inner product as the similarity metric, rather than sticking to Euclidean distance, can simplify our computation.

Aımeur et al \cite{aimeur2007quantum} also build upon this by claiming that Quantum Computing can increase the efficiency of many clustering algorithms. They use the Grover search algorithm, to create a black box that can return distance between any two points in theory. Since clustering algorithms are dominated by distance queries, a quantum distance oracle can give many distances in superposition. We will also be leveraging this superposition concept in our paper. Their results include quantum versions of divisive clustering, $k$-medians, and neighbourhood-graph construction, focusing on efficiency. This provides the motivation that if efficiencies can be increased, so can the accuracy of clustering algorithms.

Zhang and Li \cite{zhang2025improvement} in their paper identify the same weakness as the others about the classical k-means algorithm. However, they also recognize that the existing quantum k-means still have a gap in its similarity metric. They argue that fidelity only reflects the pure quantum state and ignores the classical geometric structure within the original data boundaries. This may lead to unreasonable cluster boundaries. So they introduce a hybrid distance that combines the swap test with the original classical vectors. They use two quantum variants, but the one we will discuss is Quantum k-means with angle encoding, as it achieves the highest accuracy. They use the same data set as ours and get a 0.631 accuracy, showing that this version is better than the classical one. We are implementing a similar hybrid approach, but instead of angle encoding or amplitude encoding, we are using feature maps. We will be using different feature maps and seeing which one works the best, and try to improve upon their work.

\parencite{kerenidis2019qmeans} Kerenidis and team introduce q-means as a quantum analogue of k-means. They prove that their algorithm has a faster running time than the classical k-means algorithm, whilst still having the same accuracy. Theoretically, they claim that, given access to QRAM, one can achieve an exponential time speedup. While our focus is not on an efficient algorithm, the claim is interesting
to see the power of implementing the k-means using quantum computing.

Khan et al. \cite{khan2019kmeans} says that K-means is a natural candidate for NISQ hardware if you design shallow circuits. They keep the structure of the classical k-means algorithm but swap the inner distance step with a shallow quantum subroutine. They used the same iris data set as ours. They performed it on both the simulator and the hardware. On the simulator, they were able to achieve a performance similar to the classical algorithm, and even on the hardware, they found that NISQ devices can already execute-means clustering tasks meaningfully. We share the same overall goal with the aim of achieving a quantum-enhanced k-means on NISQ hardware.  They optimize Euclidean distance, but we use a quantum kernel to change the similarity metric.

Building on their work. Bui \cite{bui2024quantumkmeans} reviews work and comments that using polar angle points only works if the points have similar radii. Otherwise, it can misjudge which centroid is the closest, so it makes 3 proposals. We will discuss the first proposal, in which they derived a new formula for the distance. They used the same shallow circuit, but they proved that this formula recovers the exact Euclidean distance even when radii differ significantly. They used the same Iris dataset, reduced 4D to 2D, and were able to get a better accuracy than Khan
et al. \cite{khan2019kmeans} 82.67, which is 88.67. Again, our goals are aligned- to try to show that quantum similarity results in better clustering. Although they try to fix the Euclidean distance, we use a quantum kernel map to compute the similarity.

Clustering is one of the most widely used tools in machine learning and data analysis, yet classical k-means struggles when the data is high-dimensional or non-linear. These limitations arise mainly because of the reliance on the Euclidean distance in the original feature space. Instead, we will utilize the inner product to determine the "similarity" between two points. Levy et al. \cite{levy2024similarity} talks about the inner product and using it as a metric for similarity. Secondly, we will be using quantum feature maps to take the data points into a higher-dimensional feature space where they might be better separable.  These papers serve as a guide explaining how quantum computing aims to offer a more efficient implementation of the k-means algorithm
The works of Khan et al.\cite{khan2019kmeans} also give a nod as their team claimed that, despite being in the NISQ era, quantum computing offers to improve the k-means algorithm, especially in terms of efficiency. Works like these that improve efficiency motivate us to improve accuracy as well.

\section{Problem Statement}
\textit{K}-means clustering is a widely used algorithm that clusters data points based on a similarity metric, known for its simplicity and efficiency. However, as discussed, it has several limitations that can result in suboptimal clustering, increased iteration counts, and high computational costs. These issues become particularly severe in high-dimensional and complex datasets. While recent approaches have improved centroid initialization, they still rely on traditional distance metrics (such as Euclidean distance), which may not effectively capture the underlying structure of the data, especially in high-dimensional spaces.

Our research aims to address these challenges by integrating quantum computing into the \textit{k}-means algorithm. We leverage the phenomenon of superposition, where a single quantum bit (or qubit) can exist in multiple states simultaneously. To achieve this, we will utilise the Iris dataset and apply quantum feature maps, then measure the similarity between data points using a quantum kernel (the inner product between quantum states). This quantum kernel will help us construct a kernel matrix, which we can then use with the classical \textit{k}-means algorithm to find clusters and evaluate the results. The quantum kernel-based approach offers a richer and more flexible similarity metric that performs better in high-dimensional spaces, providing a potential solution to the limitations of classical \textit{k}-means. By applying quantum feature maps and kernel methods, our study seeks to improve the stability, precision, and efficiency of \textit{k}-means clustering.

\section{Methods}

\subsection{Classical K-means Algorithm}
 The pseudocode for the classical k-means algorithm is provided in Algorithm 1. The pseudocode above summarizes the standard \textit{k-means} algorithm. The algorithm alternates between assigning each data point to the nearest centroid (with respect to a chosen similarity or distance measure) and recomputing each centroid as the mean of the points assigned to it, until the assignments stabilize. We replace the classical Euclidean distance with a quantum kernel based on inner products between feature-mapped quantum states. The following subsections describe how we construct these quantum feature maps, compute the corresponding kernel matrix, and plug it into this \textit{k-means} update loop.

\begin{algorithm}[H]
\caption{KMeans($X$, $k$)}
\begin{algorithmic}[1]
\State \textbf{Input:} Dataset $X = \{x_1, x_2, \dots, x_n\}$ with $n$ points in $\mathbb{R}^d$, number of clusters $k$
\State \textbf{Output:} Cluster assignments and $k$ centroids
\State Initialize $k$ centroids $\mu_1, \mu_2, \dots, \mu_k$ (e.g., randomly choose $k$ points from $X$)
\Repeat
    \For{each point $x_i$}
        \State Assign $x_i$ to the cluster with the nearest centroid:
        \State \hspace{1cm}$c_i \gets \arg\min_{j \in \{1,\dots,k\}} \|x_i - \mu_j\|^2$
    \EndFor
    \For{each cluster $j = 1$ to $k$}
        \State Update centroid $\mu_j$ as the mean of all points assigned to it:
        \State \hspace{1cm}$\mu_j \gets \frac{1}{|C_j|} \sum_{x_i \in C_j} x_i$
    \EndFor
\Until{centroids do not change (convergence)}
\State \Return cluster assignments $\{c_i\}$ and centroids $\{\mu_j\}$
\end{algorithmic}
\end{algorithm}

\subsection{Pre-processing Data}
To conduct our research, we have selected two datasets: the Iris dataset and the Breast Cancer dataset. The Iris dataset contains 150 samples, 4 numerical features, and 3 species, while the Breast Cancer dataset contains 10 features and 569 samples. Since this is unsupervised learning, we will not be using labels yet until we get to the evaluation part. The data has to be scaled in order to produce meaningful results, as most quantum maps are sensitive to raw data with a lot of variance. We will be using 4 qubits for the Iris dataset, which matches the number of features. In some places, we will use 2 qubits to encode 4 features. For the Breast Cancer dataset, we will be using 10 qubits.

\subsection{Quantum Feature Maps}
Quantum feature maps are the key to computing similarity between different data points. The feature maps used for the work in this paper are presented in Table \ref{tab:feature_maps}. We used these feature maps to construct the kernel matrix used for estimating the similarity between data points.

\begin{table}[htbp]
\caption{\textbf{Quantum feature maps used in this work.}}
\vspace{4pt}
\label{tab:feature_maps}
\centering
\footnotesize
\setlength{\tabcolsep}{6pt}
\renewcommand{\arraystretch}{1.1}

\begin{tabular}{|c|p{6cm}|}
\hline
\textbf{No.} & \textbf{Feature Map Description} \\
\hline
1 & ZZ feature map with full entanglement \\
\hline
2 & ZZ feature map with circular entanglement \\
\hline
3 & ZZ feature map with linear entanglement \\
\hline
4 & Efficient SU2 \\
\hline
5 & Z feature map \\
\hline
6 & Dense Angle Encoding (DAE) \\
\hline
7 & Angle Encoding (AE) \\
\hline
8 & Phase Encoding \\
\hline
9 & Pauli Feature Map \\
\hline
\end{tabular}
\end{table}

\begin{figure}[H]
    \centering
    \includegraphics[width=1\linewidth]{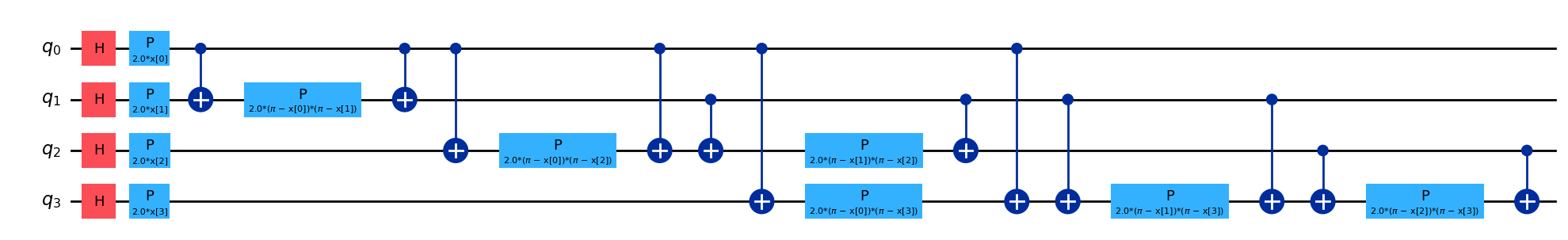}
    \caption{\textbf{ZZ feature map with linear entanglement.}}
    \label{fig_zz}
\end{figure}

\begin{figure}[H]
    \centering
    \includegraphics[width=1\linewidth]{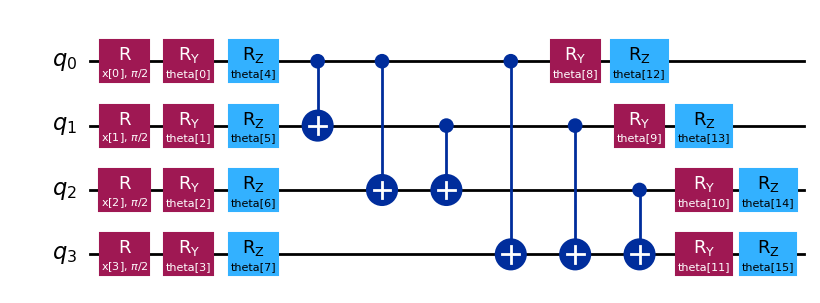}
    \caption{\textbf{Efficient SU2 feature map}.}
    \label{fig_su2}
\end{figure}

 Entanglement allows the feature map to encode correlations between different features of the dataset. Without entanglement, each qubit only carries its own feature value. With entanglement, multi-qubit interactions introduce higher-order relationships such as \(x_i x_j\) and \(x_i x_j x_k\). These higher–order terms can highlight structure that is not linearly separable in the original feature space, which may help the quantum kernel produce clearer cluster separation. ZZ and Efficient SU2 circuits using four qubits are shown in Figs. \ref{fig_zz} and \ref{fig_su2}.


\subsection{Quantum Kernel Computation}
For a given feature map $U_{\phi}(x)$, each data point is encoded as
\[
|\psi(x)\rangle = U_{\phi}(x)\,|0\rangle^{\otimes 4}.
\]
To measure the similarity between any two samples $x_i$ and $x_j$, we computed the fidelity
\[
K(x_i, x_j) = |\langle \psi(x_i) | \psi(x_j) \rangle|^2.
\]

\subsection{Quantum-assisted K-means}
The pseudocode for the quantum-assisted quantum k-means algorithm is presented in Algorithm 2. Since we have tried multiple feature maps, the only thing that changes in this algorithm is the feature map and its parameters- the rest of everything else, like initialization, cluster assignment, and centroid update, remains the same. To compute fidelity between two data points, we are using an inversion test where the probability of 0 symbolizes the similarity between 2 points. If the fidelity between two data points is 0, the corresponding states are orthogonal and have maximum distance.

\begin{algorithm}[H]
\caption{Quantum-Assisted K-Means (4D)}
\begin{algorithmic}[1]
\State \textbf{Input:} Scaled dataset $X$, feature map $U_\phi$, fixed parameters $\theta$, shots $S$
\State \textbf{Output:} Cluster labels and final centroids
\Statex
\State \textbf{Initialized Centroids:}
\Statex
\Function{QuantumSimilarity}{$x, c$}
    \State Form full parameters $(x,\theta)$ and $(c,\theta)$
    \State Prepare states $U_\phi(x)$ and $U_\phi(c)$
    \State Build overlap circuit and measure all qubits
    \State Estimate $p_0 =$ probability of outcome $0\ldots0$
    \State \Return $p_0$
\EndFunction
\Statex
\State \textbf{K-Means Loop:}
\For{$t = 1$ to $T_{\max}$}
    \For{each data point $x_i$}
        \State Compute similarities $s_{ij} = \textsc{QuantumSimilarity}(x_i, C_j)$ for $j=1..3$
        \State Assign label $y_i = \arg\max_j s_{ij}$
    \EndFor
    \For{$j = 1$ to 3}
        \State Update centroid $C_j$ as mean of points with label $j$
    \EndFor
    \If{centroids unchanged}
        \State \textbf{break}
    \EndIf
\EndFor

\State \Return labels $\{y_i\}$ and final centroids $C$
\end{algorithmic}
\end{algorithm}

\subsection{Kernel-Based Clustering}
Once the kernel matrix is computed, we will compute the distance matrix by subtracting each entry from one. If the inner product is $1$ between two states, then by this formula, it would mean that this data point belongs to that centroid. Then, compute the mean based on the clusters we have obtained for each centroid. Now repeat the entire process for the maximum iterations.

\subsection{Evaluation}
The dataset labels were then used for evaluation. Since \textit{k-means} is an unsupervised algorithm, the cluster indices produced by the algorithm (0, 1, 2) do not correspond to the true Iris species or the cancer type (0,1). To compare our predicted clusters with the ground-truth labels, we assign a class to each cluster using \emph{majority voting}.

In majority voting, we examine all data points assigned to a particular cluster and determine which true species label appears most frequently within that cluster. That label is then assigned as the representative class of the cluster. For example, if a cluster contains mostly 'Setosa' samples, that cluster is assigned to the Setosa label. This provides a method to align the unsupervised cluster outputs with the supervised ground-truth categories.
Once each cluster has an assigned label, we compute the accuracy by comparing it to the true labels.

\section{Results and Discussion}
We computed the accuracies for both the Iris and the breast cancer datasets using various feature maps, and the results are presented in Tables \ref{tab:iris_results} and \ref{tab:breast_cancer_results_short}. For the breast cancer dataset, we only computed the accuracies using the best feature maps that we found while experimenting with the Iris dataset.

\begin{table}[htbp]
\caption{\textbf{Results for the Iris dataset.}}
\vspace{4pt}
\label{tab:iris_results}
\centering
\footnotesize
\setlength{\tabcolsep}{4pt}
\renewcommand{\arraystretch}{1.15}

\resizebox{\columnwidth}{!}{%
\begin{tabular}{|p{2.3cm}|p{4.3cm}|>{\centering\arraybackslash}p{1.4cm}|>{\centering\arraybackslash}p{1.4cm}|}
\hline
\textbf{Feature Maps} & \textbf{Configuration} & \textbf{Iterations} & \textbf{Accuracy} \\
\hline
ZZFeatureMap & Linear Entanglement, 4 dim, 1 rep & 30 & 80.67\% \\
ZZFeatureMap & Circular Entanglement, 4 dim, 1 rep & 30 & 76.67\% \\
ZZFeatureMap & Full Entanglement, 4 dim, 1 rep & 30 & 88.67\% \\
Classical K-Means & Standard Scaling, 4 features & 20 & 83.33\% \\
EfficientSU2 & 4 Qubits & 30 & 89.33\% \\
EfficientSU2 & 2 Qubits & 30 & 81.33\% \\
EfficientSU2 (cc) & 2 Qubits & 30 & 48.67\% \\
\hline
\end{tabular}%
}
\end{table}

\begin{table}[htbp]
\centering
\caption{\textbf{Results for the breast cancer dataset.}}
\vspace{4pt}
\label{tab:breast_cancer_results_short}

\resizebox{0.9\columnwidth}{!}{%
\begin{tabular}{|l|l|c|}
\hline
\textbf{Feature / Model} & \textbf{Configuration} & \textbf{Accuracy} \\
\hline
EfficientSU2 & 10 Qubits & 91.39\% \\
ZZFeatureMap & Full Entanglement & 78.73\% \\
\hline
\end{tabular}%
}
\end{table}

Overall, the results indicate that integrating quantum feature maps with the \textit{k}-means clustering algorithm can achieve good performance while offering a richer similarity measure than Euclidean distance. The SU2 feature map, in particular, shows a better accuracy than the classical k-means algorithm, which produced an accuracy of $83$ percent.

The highest accuracy we were able to achieve  $88.7$ using the efficient SU2 feature map for the Iris dataset. The highest accuracy achieved for the breast cancer data set was $91.0$.We have shown the overall results of our paper in these concise tables using the different methods/maps for both datasets.

\begin{figure}[H]
    \centering
    \includegraphics[width=1\linewidth]{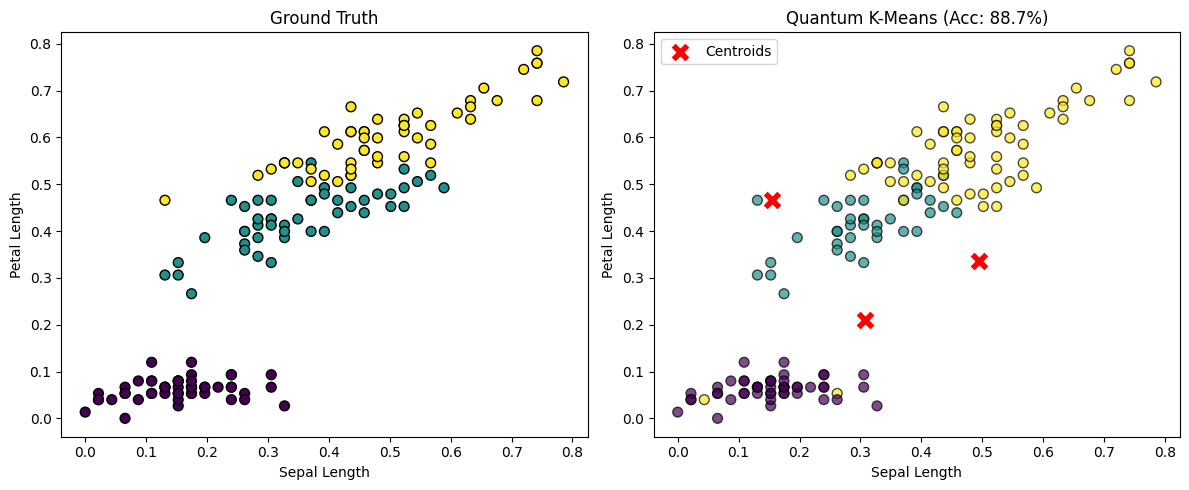}
    \caption{\textbf{Ground truth and the predicted labels for the Iris dataset using the highest-accuracy feature map.}}
    \label{fig_air}
\end{figure}
\begin{figure}[H]
    \centering
    \includegraphics[width=1\linewidth]{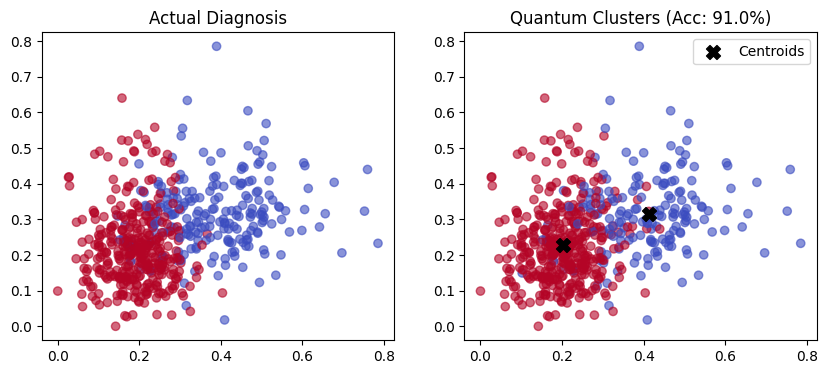}
    \caption{\textbf{Ground truth and the predicted labels for the breast cancer dataset using the highest-accuracy feature map.}}
    \label{fig_abr}
\end{figure}

The final labels and ground truth obtained for the highest-accuracy feature map are presented in Figs. \ref{fig_air} and \ref{fig_abr}. The centroids were not placed correctly during the first iterations, but with each successive iteration, the centroids were rightly placed at the center of each cluster. This demonstrates that the quantum kernel provides a meaningful similarity landscape that guides the centroids toward stable equilibrium points. The resulting cluster distribution largely follows the expected structure of the Iris dataset.

The data points with shorter sepal and petal length have been assigned Setosa, whereas data points with large petal and sepal length values have been assigned Virginica, indicating that the quantum kernel successfully captured the geometric relationships within the data.
Specifically, misclassification is occurring between Versicolor and Setosa due to the overlap evident in the box whisker plot.

Similarly, in the breast cancer dataset, although missclassification is taking place for some points, overall classification is quite accurate, indicating strong separability between the classes thanks to the quantum feature map.

\begin{figure}[H]
    \centering
    \includegraphics[width=0.8\linewidth]{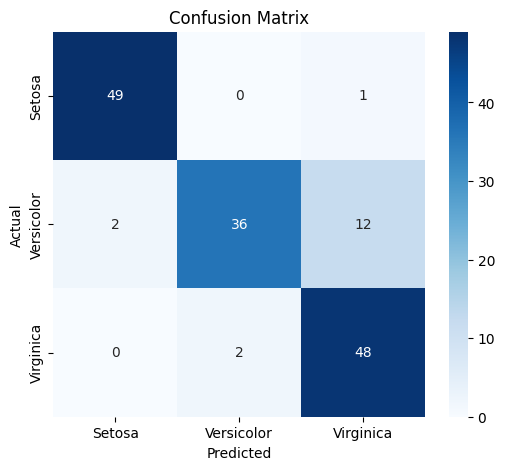}
    \caption{\textbf{Confusion matrix for the Iris dataset using the highest-accuracy feature map.}}
    \label{fig1}
\end{figure}

\begin{figure}[H]
    \centering
    \includegraphics[width=0.8\linewidth]{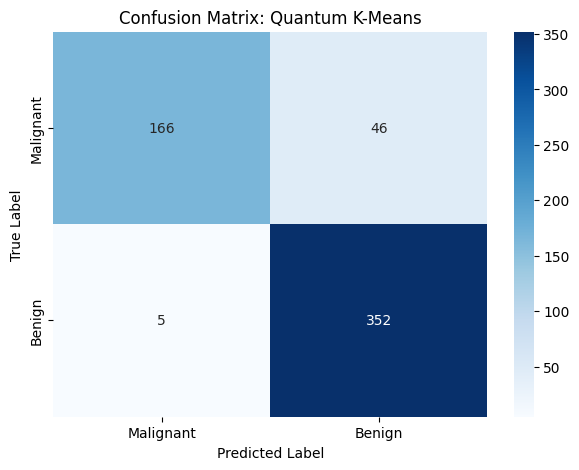}
    \caption{\textbf{Confusion matrix for the breast cancer dataset using the highest-accuracy feature map.}}
    \label{fig2}
\end{figure}

The confusion matrices for both datasets are given in Figs. \ref{fig1} and \ref{fig2}.

We also compared the clustering accuracy with quantum k-means with baseline classical k-means. As can be seen from Tables \ref{tab:iris_classical_quantum} and \ref{tab:breast_cancer_classical_quantum}, quantum k-means provide better accuracies than classical baseline k-means.

\begin{table}[htbp]
\centering
\caption{\textbf{Best quantum feature maps vs classical k-means accuracies (Iris dataset).}}
\vspace{4pt}
\label{tab:iris_classical_quantum}
\footnotesize
\setlength{\tabcolsep}{4pt}
\renewcommand{\arraystretch}{1.15}

\resizebox{1.0\columnwidth}{!}{%
\begin{tabular}{|l|p{4.2cm}|c|}
\hline
\textbf{Feature / Model} & \textbf{Configuration} & \textbf{Accuracy} \\
\hline
EfficientSU2 & 4 Qubits & 88.67\% \\
ZZFeatureMap & Full Entanglement & 88.00\% \\
Classical K-Means & Standard Scaling, 4 features & 83.33\% \\
\hline
\end{tabular}
}
\end{table}

\begin{table}[htbp]
\centering
\caption{\textbf{Best quantum feature maps vs classical k-means accuracies (breast cancer dataset).}}
\vspace{4pt}
\label{tab:breast_cancer_classical_quantum}
\footnotesize
\setlength{\tabcolsep}{4pt}
\renewcommand{\arraystretch}{1.15}

\begin{tabular}{|l|p{3cm}|c|}
\hline
\textbf{Feature / Model} & \textbf{Configuration} & \textbf{Accuracy} \\
\hline
EfficientSU2 & 10 Qubits & 91.39\% \\
Classical K-Means & Standard Scaling, 10 features & 89.98\% \\
ZZFeatureMap & Full Entanglement & 81.90\% \\
\hline
\end{tabular}
\end{table}

\begin{table}[h]
\centering
\caption{\textbf{Clustering Metrics for Breast Cancer Dataset}}
\vspace{4pt}
\begin{tabular}{|l|c|c|}
\hline
\textbf{Feature / Model} & \textbf{ARI} & \textbf{AMI} \\
\hline
EfficientSU2 & 0.6407 & 0.5514 \\
Classical K-Means & 0.6348 & 0.5512 \\
ZZFeatureMap & 0.3478 & 0.2511 \\
\hline
\end{tabular}
\end{table}

\begin{table}[h]
\centering
\caption{\textbf{Clustering Metrics for Iris Dataset}}
\vspace{4pt}
\begin{tabular}{|l|c|c|}
\hline
\textbf{Feature / Model} & \textbf{ARI} & \textbf{AMI} \\
\hline
EfficientSU2 & 0.7302 & 0.7551 \\
Classical K-Means & 0.7163 & 0.7387 \\
ZZFeatureMap & 0.6867 & 0.7010 \\
\hline
\end{tabular}
\end{table}

\section{Conclusions}
In this work, we explored a quantum-enhanced approach to the classical \textit{k-means} clustering algorithm by replacing the traditional Euclidean distance with a quantum kernel derived from the inner product of feature-mapped quantum states. Our study demonstrates that quantum feature maps—especially those utilising entanglement—offer a richer and more flexible similarity metric that can capture nonlinear structures more effectively than classical distance-based methods. While misclassifications persist, particularly in regions where the original data classes overlap, the overall performance supports the idea that quantum kernels can improve cluster quality without altering the core structure of the algorithm. Here is a comparison between the quantum vs classical algorithms for both datasets.

We also included the ARI and AMI scores for both data sets. Table VI compares classical and Quantum algorithms for the Breast Cancer dataset, whereas Table VII compares them for the Iris dataset. The table shows that the ARI and AMI scores of the EfficientSU2 feature map are better than those of classical K-means in both datasets, indicating that the quantum algorithms perform better.

This study highlights important insights- that, despite being in the NISQ era, quantum computing can be integrated in classical machine learning, showing that it can produce better results on higher-dimensional datasets. Future work may include testing deeper or adaptive feature maps on better hardware, producing even more accurate results. Together, these directions point toward the growing potential of quantum machine learning to provide both accuracy and stability benefits for clustering tasks in the coming years.

\bibliographystyle{plain}

\bibliography{refs}

\end{document}